# Colossal magnetoresistance in the insulating ferromagnetic double perovskites Tl$_2$NiMnO$_6$: A neutron diffraction study

Lei Ding[*a], Dmitry D. Khalyavin[a], Pascal Manuel[a], Joseph Blake[b], Fabio Orlandi[a], Wei Yi[c], Alexei A. Belik[†d]

[a] ISIS Facility, Rutherford Appleton Laboratory, Harwell Oxford, Didcot OX11 0QX, United Kingdom.

[b] Department of Physics, Royal Holloway, University of London, Egham, TW20 OEX, United Kingdom

[c] International Center for Materials Nanoarchitectonics (WPI-MANA), National Institute for Materials Science (NIMS), Namiki 1-1, Tsukuba, Ibaraki 305-0044, Japan.

[d] Research Center for Functional Materials, National Institute for Materials Science (NIMS), Namiki 1-1, Tsukuba, Ibaraki 305-0044, Japan.

**Abstract**

In the family of double perovskites, colossal magnetoresistance (CMR) has been so far observed only in *half-metallic ferrimagnets* such as the known case Sr$_2$FeMoO$_6$ where it has been assigned to the tunneling MR at grain boundaries due to the half-metallic nature. Here we report a new material−Tl$_2$NiMnO$_6$, a relatively ordered double perovskite stablized by the high pressure and high temperature synthesis−showing CMR in the vicinity of its Curie temperature. We explain the origin of such effect with neutron diffraction experiment and electronic structure calculations that reveal the material is a *ferromagnetic insulator*. Hence the ordered Tl$_2$NiMnO$_6$ (~70% of Ni$^{2+}$/Mn$^{4+}$ cation ordering) represents the first realization of a *ferromagnetic* insulating double perovskite, showing CMR. The study of the relationship between structure and magnetic properties allows us to clarify the nature of spin glass behaviour in the disordered Tl$_2$NiMnO$_6$ (~31% of cation ordering), which is related to the clustering of antisite defects and associated with the short-range spin correlations. Our results highlight the key role of the cation ordering in establishing the long range magnetic ground state and lay out new avenues to exploit advanced magnetic materials in double perovskites.

---

[*] Corresponding author: E-mail: lei.ding.ld@outlook.com (L. Ding)
[†] alexei.belik@nims.go.jp (A. A. Belik)

## 1. Introduction

Perovskites, a fertile playground for the study of strong interplay between spin, orbital and charge degrees of freedom, can accommodate a large variety of cation species due to their flexible crystal structure, giving rise to diverse and fascinating electronic and magnetic properties [1-7]. In ordered double perovskites, with chemical formula $A_2BB'O_6$ (A= rare earth or alkaline, B/B'= transition metal), B and B' cations generally form either rock-salt or layered type superstructure. The former, which leads to alternate stacking of $BO_6/B'O_6$ octahedra, is the most common type to minimize the electrostatic energy arising from the different charges of the two cations, and/or the elastic energy due to their different ionic sizes [8-16]. When the radius or the charge of B and B' cations is similar, the probability of antisite (AS) defects (partial occupation of B cations in B' sites and vice versa) increases up to a fully disordered state, effectively tuning the macroscopic properties of the system [11].

One of the most intriguing properties of double perovskites is the colossal magnetoresistance (CMR), typically arising from the half-metallic nature of their electronic structures [17-19]. For example, half-metallic $Sr_2FeMoO_6$ shows a CMR in low magnetic field associated with a ferrimagnetic order above room temperature, providing a promising system for technological applications [17].

A rather small MR with a value of 3% associated with a ferromagnetic (FM) order at 200 K was also identified in semiconducting La$_2$NiMnO$_6$ [20-21]. However, no CMR has been found *hitherto,* to the best of our knowledge, in insulating FM double perovskites. In addition, AS disorder has been shown to have a dramatic influence on the magnetic and magneto-transport properties, as for example in Sr$_2$FeMoO$_6$ where it affects the half-metallic character and lowers the tunneling MR effect [22-24]. Its influence is also pronounced in La$_2$NiMnO$_6$ where the MR effect is practically destroyed by the presence of just 20% of AS defects [25]. Therefore, it is critical to understand how synthetic conditions or post-synthesis treatments can influence the cationic ordering, and how this, in turn, controls the macroscopic properties of double perovskite systems. Herein, we present a new double perovskite Tl$_2$NiMnO$_6$ that contains small A-site cation with different degrees of Ni$^{2+}$/Mn$^{4+}$ AS disorder, depending on the synthesis procedure. This provides a case study of the effect of cation ordering on the magnetic properties of insulating double perovskites. We found that the nominally ordered sample exhibits a colossal low-field MR coupled to a long-range FM ordering, while only short-range spin correlations are present in the disordered compound.

**2. Experimental and theoretical calculations**

*2.1 Materials synthesis*

Two polycrystalline samples of Tl$_2$NiMnO$_6$ with different degrees of the B-site cation ordering were prepared under high pressure and high temperature (HPHT) conditions in a belt-type apparatus. The stoichiometric mixture of reagent-grade Tl$_2$O$_3$, NiO and MnO$_2$ was placed in a Au capsule and then treated at 6 GPa and about 1570 K (ordered sample) or about 1470 K (disordered sample) for 2 h (the duration of heating to the synthesis temperatures was 10 min). After the heat treatment, the samples were quenched to room temperature, and the pressure was slowly released. Caution! Due to high toxicity of the Tl-containing compounds a special precaution should be added at preparation and handling of this double perovskite.

*2.2 Magnetic, heat capacity and magneto-transport measurements*

Magnetic susceptibilities were measured using SQUID magnetometers [Quantum Design, MPMS-7T(XL) and MPMS-1T] between 2 and 350 K under magnetic fields of 3 Oe, 100 Oe and 1 T with both zero-field-cooled (ZFC) and field-cooled (FC) conditions. Magnetic hysteresis loops were measured at 2, 10, 20, 50, 60 and 100 K using MPMS-7T from -7 T to 7 T. Ac susceptibility measurements were performed using a Quantum Design MPMS-1T instrument from 2 to 110 K at frequencies of 2, 110, 300 and 500 Hz and an applied oscillating magnetic field of 5 Oe.

Resistivity and magnetoresistance measurements were performed on the ordered sample using both ac impedance spectroscopy (in the frequency range of 0.1 Hz-2 MHz) and standard four-probe method. For these measurements, pellets (5.5 mm in diameter and 2 mm thickness) were cut into squares with the size of about 4 mm x 4 mm and polished to the thickness of about 1 mm. Contacts were made as lines on the pellet surface with the 4 mm length. A commercial Physical Property Measurement System (PPMS) was used for the resistivity measurements. The maximum current was limited by 1 microA, and the current was automatically adjusted by the PPMS to make measurements at low temperatures (at 140 K, the measurement current was about 0.03 microA). For the disordered sample, in order to produce a dense ceramic suitable for the heat capacity and magnetic-transport measurements, the as-prepared disordered sample was pelletized under a high pressure of 6 GPa at room temperature. The heat capacity measurements were carried out using a relaxation technique with a Quantum Design PPMS in the temperature range of 2-200 K on cooling. The pelletized samples were mounted on a sample platform with Apiezon N-grease for better thermal contact.

*2.3 Neutron powder diffraction and data analysis*

Neutron powder diffraction (NPD) experiments were carried out at the ISIS pulsed neutron and muon facility of the Rutherford Appleton Laboratory (UK), on the WISH diffractometer [26] located at the second target station. Powder samples (0.9 g for the ordered sample and 0.4 g for the disordered sample) were loaded into cylindrical vanadium cans with 3 mm in diameter and measured in the temperature range of 1.5 - 120 K using an Oxford Instrument Cryostat. The neutron diffraction data were collected at 1.5 and 120 K each for 30 minutes, and short exposions with 10 minutes at 40, 60, 80 and 100 K were recorded. Rietveld refinements of the crystal and magnetic structures were performed using the Fullprof program [27] against the data measured in detector banks at average 2θ values of 58°, 90°, 122° and 154°, each covering 32° of the scattering plane. Group theoretical calculations were done using ISODISTORT [28] and Bilbao Crystallographic Server (Magnetic Symmetry and Applications) software [29].

*2.4 Computational Method*

Spin-polarised σ-DFT+U calculations [30-31], pertaining to understanding the electronic structure of $Tl_2NiMnO_6$ were performed using CASTEP (v17.2) [32] software, which exploits a plane-wave basis set with ultrasoft-pseudopotentials [33-34] and solves the Kohn-Sham equations [35-36]. The electronic exchange and correlation was described using spin-polarised generalised gradient approximation (σ-GGA) expressed by Perdew, Burke and Ernzerhof [37]. The plane-wave cut-off energy was determined at 800eV within the convergence criterion of 0.01GPa and CASTEP's inbuilt on-the-fly generator (C17) constructed the ultrasoft-pseudopotentials. The Monkhorst-Pack K-point grid for sampling the Brillouin zone [38] was converged at 11-11-9 with the same criterion as the cut-off energy. Use of σ-DFT+U, as described by Cococcioni et al [30] and implemented in CASTEP was applied to the 3d orbitals of the Mn and Ni ions. The U values were tested in the range of 0 - 5 eV, with steps of 0.25 eV. This was graphically compared to the stress on the unit cell (GPa). On the U vs stress graph, a plateau formed (1.25-1.50eV for Mn and 1.00-1.25eV for Ni) which offered the values that the U parameter takes. Values from the plateau were tested to see the impact that they had on the band structure. Through this test, the effective values of U for Mn and Ni were suggested to be 1.50 eV and 1.25 eV, respectively.

## 3. Results and discussion

*3.1 Crystal structure and cation disorder*

The crystal structures of the synthetic samples were determined by neutron diffraction collected at 120 K. Several structural models were tested in the refinement procedure, and the best resolution shows that both $Tl_2NiMnO_6$ samples crystallize with monoclinic $P2_1/n1'$ space group (No.14, cell choice 2), isostructural to $In_2NiMnO_6$ [39-40]. Rietveld plots of the neutron diffraction patterns for both samples are shown in Figure 1a,b, and the refined structural parameters are summarized in Table S1 in the supporting information. The quantitative refinement yields the occupancy for Ni/Mn sites to be 0.85(1)/0.15(1) and 0.658(2)/0.342(2), resulting in 70% and 31.6% of cation ordering, in the nominally ordered and disordered samples, respectively. We quantify the ordering degree as $|P_{Ni}-P_{Mn}|/(P_{Ni}+P_{Mn})*100\%$, where $P_{Ni}$ and $P_{Mn}$ denote the probabilities of occupying B(B')-site by Ni and Mn. The concentration of AS defects in this system appears to be controlled mostly by the synthesis temperature. Indeed, an increase of the latter by 100 K, leaving all the other synthesis conditions unchanged, gives sufficient diffusion energy to the B/B' species to order themselves in the double perovskite structure (see the experimental section for detailed description of the sample preparation). The crystal structure of ordered $Tl_2NiMnO_6$ is presented in Figure 1c. It consists of corner-sharing $MnO_6$ and $NiO_6$ octahedra arranged in the typical double perovskite rock-salt structure. The B-O bond distances of the two octahedral sites as well as bond valence sum calculations are presented in Table S3. They confirm the $Mn^{4+}$ and $Ni^{2+}$ charge distribution. Due to the small size of the $Tl^{3+}$ cation, significant in-phase and out-of-phase octahedral tilting with the $a-a-c+$ pattern in the Glazer notation [41] occurs (Figure 1d). The average Ni-O-Mn bond angle in the ordered $Tl_2NiMnO_6$ is 144.2°, which is larger than that in $In_2NiMnO_6$ (140.4°), where a spiral antiferromagnetic (AFM) structure is disclosed [40]. Another double perovskite $Lu_2NiMnO_6$, with the same monoclinic symmetry and an average bond angle of 143.5°, exhibits a FM transition at 40 K [42]. These observations highlight a crossover between FM and AFM ground states around a bond angle of 140°. Generally, based on the Goodenough-Kanamori rules [43-44], the 180° super-exchange interaction between half-filled $e_g$ orbital of $Ni^{2+}$ ($3d^8$, $t_{2g}^6 e_g^2$) and empty $e_g$ orbital of $Mn^{4+}$ ($3d^3$, $t_{2g}^3 e_g^0$) is expected to be FM, as revealed in most $A_2NiMnO_6$ systems [11]. The decrease of the bond angle, however, reduces the strength of this super-exchange and facilitates the role of AFM 90° and next-nearest-neighbor interactions as indicated by the monotonic suppression of the FM transition temperature till the crossover to AFM phases takes place around 140°.

The bond lengths and angles of the disordered sample (Table S3) are very similar to those of the ordered one. Nevertheless, it is interesting to point out the presence of nuclear diffuse scattering around the ½½½-type reflections, referring to the pseudo-cubic perovskite subcell (d-spacing~4.4 Å) (see Figure S1 in the supporting information). These reflections are characteristic of the rock-salt type of cation ordering in the B-site of perovskite lattice. The observation of Bragg reflections and diffuse scattering around this position, therefore, indicates a coexistence of long- and short-range cation orderings. This, in turn, points to a tendency of the AS defects to form clusters (regions with high concentration of AS defects). Such a tendency is also observed in $A_2FeMoO_6$ (A=$Sr^{2+}$, $Pb^{2+}$) compounds in which the clustering of AS defects was proved using local probes [15, 45]. As it will be shown below this in fact has a significant influence on the magnetic properties of the system.

*3.2 Magnetic properties*

The temperature-dependent dc magnetic susceptibilities of ordered and disordered $Tl_2NiMnO_6$ are shown in Figure 2a and b. A rapid deviation of the curves from Curie-Weiss behaviour occurs around $T_C$=80 K for the ordered and 70 K for the disordered sample. The zero-field cooling (ZFC) and field cooling (FC) curves of the ordered sample apparently bifurcate at $T_C$ followed by a broad maximum centered at around 60 K in ZFC. This behavior is absent in the magnetic susceptibilities measured under a magnetic field of 1T, as shown in Figure S2. In order to further characterize the magnetic phase transition, we measured the specific heat of ordered $Tl_2NiMnO_6$, shown in Figure S4. A small cusp at 80 K is indicative of the ferromagnetic transitions, in good agreement with our magnetic susceptibility data.

Similar irreversibility is also present in the disordered sample under a small magnetic field of 100 Oe; however, a small kink is observed around 15 K, probably related to the high concentration of AS defects as recently suggested for partially disordered $La_2NiMnO_6$ [25, 46]. As shown in Figure S4, the temperature-dependent specific heat of disordered $Tl_2NiMnO_6$ shows no anomaly in the whole measured temperature range, indicating the absence of magnetic phase transition. This also confirms that the small kink shown in the dc magnetic susceptibility curve does not stem from any phase transition but the concentration of AS defects. The Curie-Weiss law has been employed to fit the magnetic susceptibility in the high temperature range. The fit yields Weiss temperatures $\theta$=98 K and 77 K, and effective magnetic moments $\mu_{eff.}$=5.04 $\mu_B$/f.u. and 5.11 $\mu_B$/f.u. for the ordered and disordered sample, respectively. The positive $\theta$s are close to $T_C$ and indicate an average FM exchange with negligible frustration.

To gain a further insight into the nature of the magnetism in $Tl_2NiMnO_6$, frequency-dependent ac magnetic susceptibility was measured under an oscillating magnetic field of 5 Oe. As shown in Figure 2c and S3, in the ordered sample, both the real $\chi'$ and imaginary $\chi''$ parts of ac susceptibility consistently show a sharp and slightly frequency-dependent peak at $T_C$. However, around 60 K (glassy temperature $T_g$), a broad maximum with substantial frequency dependence is observed, suggesting a presence of spin-glass like component. This behaviour is also seen in the disordered sample, where only a broad, strongly frequency-dependent maximum is seen around the same temperature (Figure 2d and the inset). This broad feature is thus likely related to the glassy phase, whose fraction clearly correlates with the concentration of AS defects present in both compounds. As shown by the neutron diffraction data, the AS defects in the nominally disordered sample tend to aggregate, creating small regions with only short range $Ni^{2+}/Mn^{4+}$ cation ordering. These regions are probably responsible for the spin-glass like behavior. For spin glass systems, a quantitative measure of the frequency shift is given by $\nu=\Delta T_g/(T_g\Delta\log(f))$ [47], where $T_g$ denotes the maximum in the ac susceptibility. We obtained $\nu=0.0117$ and $0.0066$ for the ordered and disordered sample, respectively. Both values are within the range of 0.0045-0.08 typical for canonical spin glasses [47]. It is interesting to note that the same frequency-dependent behaviour has also been demonstrated in partially ordered $La_2NiMnO_6$ [25] where it was assigned to a re-entrant spin glass phase in a homogenous magnetic state. Instead, our neutron diffraction data and the comparison between the ordered and disordered samples allow us to conclude that the spin glass behaviour is likely to be related to the clustering of AS defects and the associated short range magnetic ordering.

Figure 3 shows the magnetic hysteresis loops of the ordered and disordered $Tl_2NiMnO_6$ samples at various temperatures. The former is practically linear above 1 T at 2 K, yielding the spontaneous magnetic moment ~3.0 $\mu_B$/f.u. and the coercive field about 190 Oe (a soft magnet behaviour, see in Figure S3c). The moment is substantially smaller than 5 $\mu_B$/f.u. expected for a perfectly ordered sample. The reduction in magnetic moment can be attributed to the presence of AS disorder. Indeed, the Ni-O-Ni and Mn-O-Mn magnetic paths, generated by the AS defects, are expected to be AFM, thus acting against the Ni-O-Mn ferromagnetic exchange, thereby reducing the saturated moment. In fact, a reduced ordered magnetic moment has also been found in slightly disordered double perovskite $Lu_2NiMnO_6$ [42] and heavily disordered $Sr_2FeMoO_6$ with only

18% of the cation ordering [24]. The magnetization of the nominally disordered Tl$_2$NiMnO$_6$ sample is substantially smaller than that of the ordered one and does not vary linearly even in high magnetic fields, which is typical for spin glasses.

*3.3 Neutron magnetic diffraction analysis*

The neutron powder diffraction (NPD) data collected from 1.5 K to 120 K on the ordered sample reveal a clear intensity increase of the 110 and 002 nuclear reflections below T$_C$ (Figure 4a), indicating the onset of the long range FM ordering. However, NPD data show no further magnetic phase transition around 60 K, probably implying a local character for the anomalies observed in the ac susceptibility data. All the magnetic reflections can be indexed by ***k*=0** propagation vector. It is reasonable to impose that the two sublattices made up of Ni cations and Mn cations, respectively, order coincidently because of the strong Ni-O-Mn superexchange interactions. Representation analysis applied to the parent space group *P2$_1$/n1'* and the propagation vector corresponding to the Gamma point of the Brillouin zone indicates two active irreducible representations (*mGM1+* and *mGM2+*) for 2d and 2c magnetic atom positions. The irreducible representation *mGM1+*, allowing ferromagnetic order along the b-axis, corresponds to *P2$_1$/n* subgroup while the *mGM2+*, transforming the two atoms ferromagnetically along the a- and c-axes, is associated with *P2$_1$'/n'* subgroup. We have tested these two magnetic models against neutron diffraction data. Quantitative Rietveld refinements revealed that the ordered FM moments are aligned along the b-axis, represented by *P2$_1$/n* magnetic symmetry. It is worth pointing out that the spin structure transformed by *mGM1+* allows antiferromagnetic components along the a- and c-axes, but these AFM modes are secondary order parameters and are beyond the resolution limit of the present neutron diffraction experiment. The final refinement of the neutron diffraction pattern collected at 1.5 K and a schematic representation of the magnetic structure are shown in Figure 4b and c, respectively. The refined magnetic moments at 1.5 K are 1.51(4) $\mu_B$/Mn$^{4+}$ and 0.99(5) $\mu_B$/Ni$^{2+}$, yielding the total ferromagnetic moment of 2.50(5) $\mu_B$/f.u.

In contrast to the ordered sample, no magnetic Bragg contribution to the scattering can be found for the disordered sample down to the lowest measured temperature 1.5 K, indicating the absence of a long-range ordered component. This also rules out any phase transition around 15 K (the anomaly observed in the magnetic susceptibility under 100Oe). A careful analysis of the diffraction pattern measured at the forward scattering detector bank disclosed an increase of the low momentum scattering Q→0 below 60 K, as shown in Figure 5a, suggesting a formation of short range FM spin correlations. The lack of long range magnetic ordering in this sample clearly correlates with the high concentration of the AS defects. Moreover, a weak diffuse scattering appears around the 110 and 002 reflections at low temperatures (Figure 5b). This scattering can be attributed to the Mn-O-Mn and Ni-O-Ni AFM interactions in the AS-rich regions of the sample, where these interactions dominate.

*3.4 Magnetoresistance*

The MR properties of the ordered Tl$_2$NiMnO$_6$ sample were investigated by measuring its resistivity under various magnetic fields through both ac impedance spectroscopy and the standard four-probe method. The latter, however, failed below 140 K due to the high resistivity of the sample. The former method allowed us to evaluate the MR properties of the sample down to 70 K as illustrated in Figure 6. It should be pointed out that both methods delivered quantitatively consistent results in the temperature range above 140 K (see Figure S5 and S6 in the supporting information). As shown in Figure 4a, ordered

Tl$_2$NiMnO$_6$ was found to be insulating featured by a high resistivity 5×10$^6$ Ω cm at 140 K.

The isothermal MR curves for the ordered Tl$_2$NiMnO$_6$ measured in the temperature range of 80 - 100 K reveal a considerable MR effect. More interestingly, below T$_C$, the field dependence of resistivity exhibits a sharp drop in the low-field regime, being nearly inversely proportional to temperature. MR reaches its maximum value of 80% at 70 K. Below this temperature, it was not possible to measure the resistivity in zero field even with the impedance method, due to the considerable insulating character of the material. In spite of the fact that we do not have MR data below 70 K, a comparison between the variation of MR at 70 K and the field-dependent magnetization at 50 K (the inset of Figure 3a) enables us to gain more insight into the origin of this interesting phenomenon. As mentioned above, the FM component practically saturates around 1 T (Figure. 3a) and the value of MR reaches 45% in this field (Figure 6). Thus, the steep drop in resistivity clearly correlates with the sharp increase in magnetization. This signals that the origin of the negative CMR might be related to a suppression of spin-dependent scattering. This conclusion has a further support from σ-DFT+U calculations of the electronic structure of Tl$_2$NiMnO$_6$. In contrast to the ordered sample, disordered Tl$_2$NiMnO$_6$ shows a tiny MR effect which enhances slightly with increasing magnetic field, as shown in Figure S7. At 50 K, the MR reaches its maximum value of 7 % under a magnetic field of 8 T, significantly smaller than that of the ordered sample. These observations imply that it is the long-range FM ordering that triggers the CMR effect in the ordered Tl$_2$NiMnO$_6$.

*3.5 DFT calculations*

The calculations have been done for a hypothetical case of fully ordered Mn and Ni cations, using generalized gradient approximation (GGA+U) to account for electronic correlations. A set of test runs with the coulomb correction U varying in the range of 0 - 5 eV have been performed with the final choice of this parameter to be 1.50 and 1.25 eV for Mn and Ni respectively (see supporting information for details). The calculations confirm that the magnetic ground state of Tl$_2$NiMnO$_6$ is FM with the spins polarized along the *b-axis*, in agreement with the experimentally found magnetic space group *P2$_1$/n*. The Ni$^{2+}$ and Mn$^{4+}$ cations possess magnetic moments of 1.45 µ$_B$ and 3.24 µ$_B$ respectively. The corresponding spin-polarized density of states (σ-DOS), projected density of states (PDOS) [48] and the electronic band structure are shown in Figure 7 (see also Figure S8 in the supporting information). The PDOS indicates that there is a complete domination of Ni and Mn α- (spin up) and β- (spin down) electrons below the Fermi Energy, in favour of the α-electrons, thus accounting for the FM state. The calculations also reveal the presence of clear band gaps of magnitude ~0.2 eV and ~0.7 eV for the α- and β-electrons, respectively. Although, these values cannot be taken as quantitative characteristics of the material, due to the semi-empirical choice of the U parameter, qualitatively, the difference in the band gaps for electrons with different spin polarization was reproduced in the calculations using other values of U.

Large MR in low-field has previously been found in several double perovskites such as Sr$_2$FeMoO$_6$ and Mn$_2$FeReO$_6$ where the origin of this effect was interpreted in terms of tunneling MR at grain boundaries due to their half-metallic nature [17-19]. The electronic structure of Tl$_2$NiMnO$_6$ has a pronounced insulating character, however, the difference in band-gaps for the spin up and down channels, implies that at finite temperature there is unbalance between the α- and β-electrons in the conduction band. This results in the presence of effectively polarized charge carriers and dependence of the transport properties on applied magnetic field. The magnetization data presented above revealed that $Tl_2NiMnO_6$ shows the characteristics of a soft FM. This implies that in the presence of a low magnetic field, the magnetic domains can be readily aligned along the field direction, thus effectively decreasing the spin-dependent scattering between the domains. We therefore suggest that the CMR in $Tl_2NiMnO_6$ is attributed to a suppression of spin-dependent scattering of the polarized carriers. For many years, large MR was considered to be a direct consequence of half-metallicity in *ferrimagnetic* double perovskites [11]. The finding of low-field CMR in a *ferromagnetic* insulator thus provides new insights into the development and design of novel magnetic functional materials in double perovskites.

## 4. Conclusions

In conclusion, a new metastable double perovskite $Tl_2NiMnO_6$ was prepared with different degrees of the B-site cation ordering by varying the high-pressure, high-temperature synthesis conditions. The cation ordering is the key factor controlling the magnetic ground state of this material. By changing the cation ordering degree from 70% to 31%, one can suppress the long-range FM order, resulting in short range spin correlations and glass-like behaviour. The ordered ferromagnetic $Tl_2NiMnO_6$ exhibits a low-field CMR effect reaching 80% at 70 K, which is ascribed to a suppression of spin-dependent scattering between ferromagnetic domains. In contrast to *ferrimagnetic* half-metallic double perovskites, where CMR is typically found, ordered $Tl_2NiMnO_6$ represents the first realization of a *ferromagnetic* insulating double perovskite, showing CMR. Our results open new avenues for material design for functional magnets in the family of double perovskites.

**Acknowledgements**

L.D. thanks support from the Rutherford International Fellowship Programme (RIFP). This project has received funding from the European Union's Horizon 2020 research and innovation programme under the Marie Skłodowska-Curie Grant Agreements No. 665593 awarded to the Science and Technology Facilities Council. D.D.K. and P.M. acknowledge TUMOCS project. This project has received funding from the European Union's Horizon 2020 research and innovation programme under the Marie Skłodowska-Curie Grant Agreements No. 645660. We are grateful to Keith Refson for informative discussions and invaluable advice.

**Figure:**

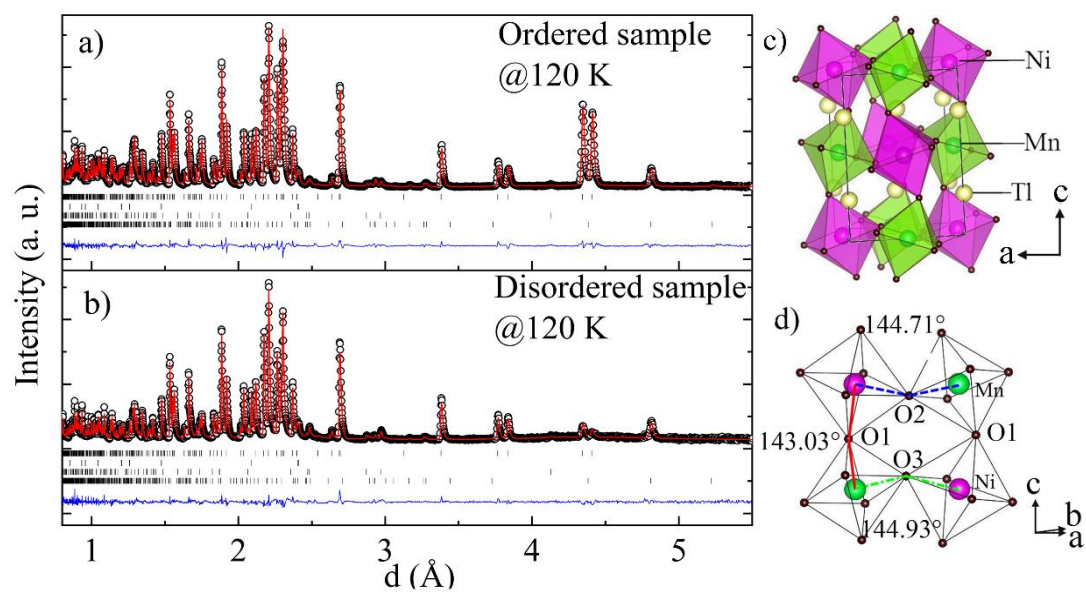

Figure 1. a)-b) Rietveld refinements of the neutron powder diffraction patterns of the ordered and disordered samples collected at 120 K. The nuclear reflections of $Tl_2NiMnO_6$ are denoted by upper tick marks. The reflections marked in the second, third and fourth lines correspond to impurity phases NiO (1.81(3) wt% and 1.21(3) wt %), $Tl_2O_3$ (1.8(1) wt % and 3.3(2) wt %) and $Tl_2CO_3$ (0.32(7) wt % and 0.4(1) wt %), respectively. c)-d) The crystal structure of the ordered $Tl_2NiMnO_6$ perovskite where the relevant bond angles Ni-O-Mn and exchange interaction paths are marked

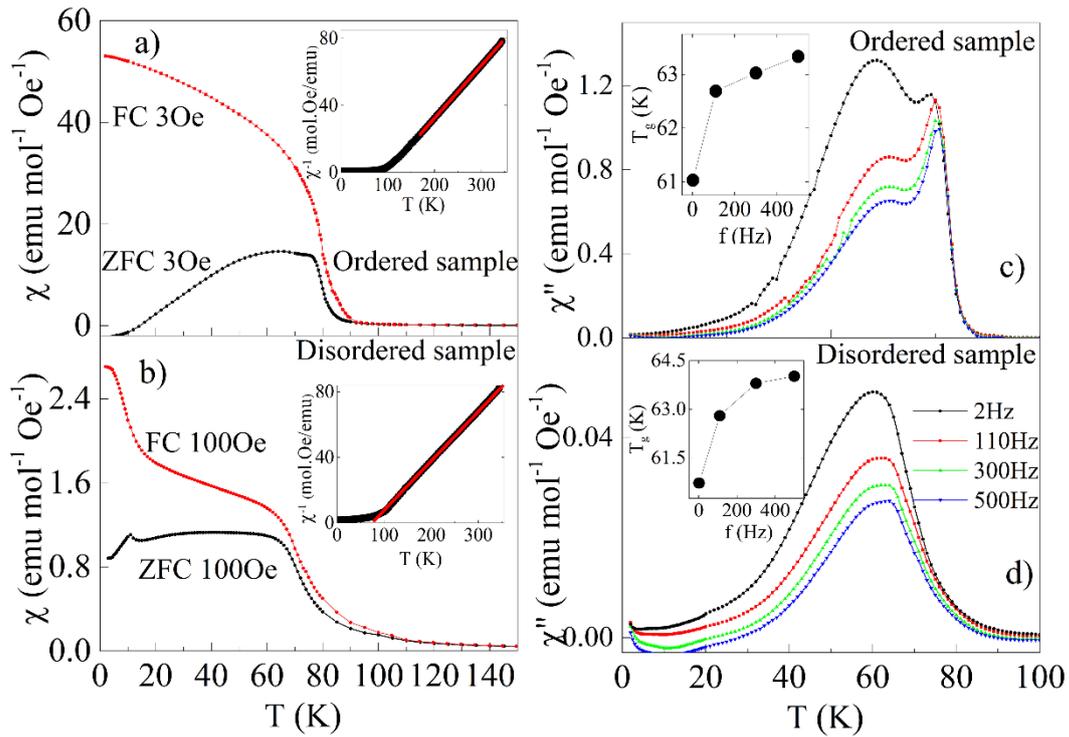

Figure 2. Temperature dependence of the dc magnetic susceptibility for the ordered a) and disordered b) $Tl_2NiMnO_6$ samples. Insets show the Curie-Weiss fittings. The imaginary parts of the ac magnetic susceptibility at various frequencies for the ordered c) and disordered d) samples under an oscillating magnetic field of 5 Oe. Insets demonstrate the glassy temperature $T_g$ as a function of frequency

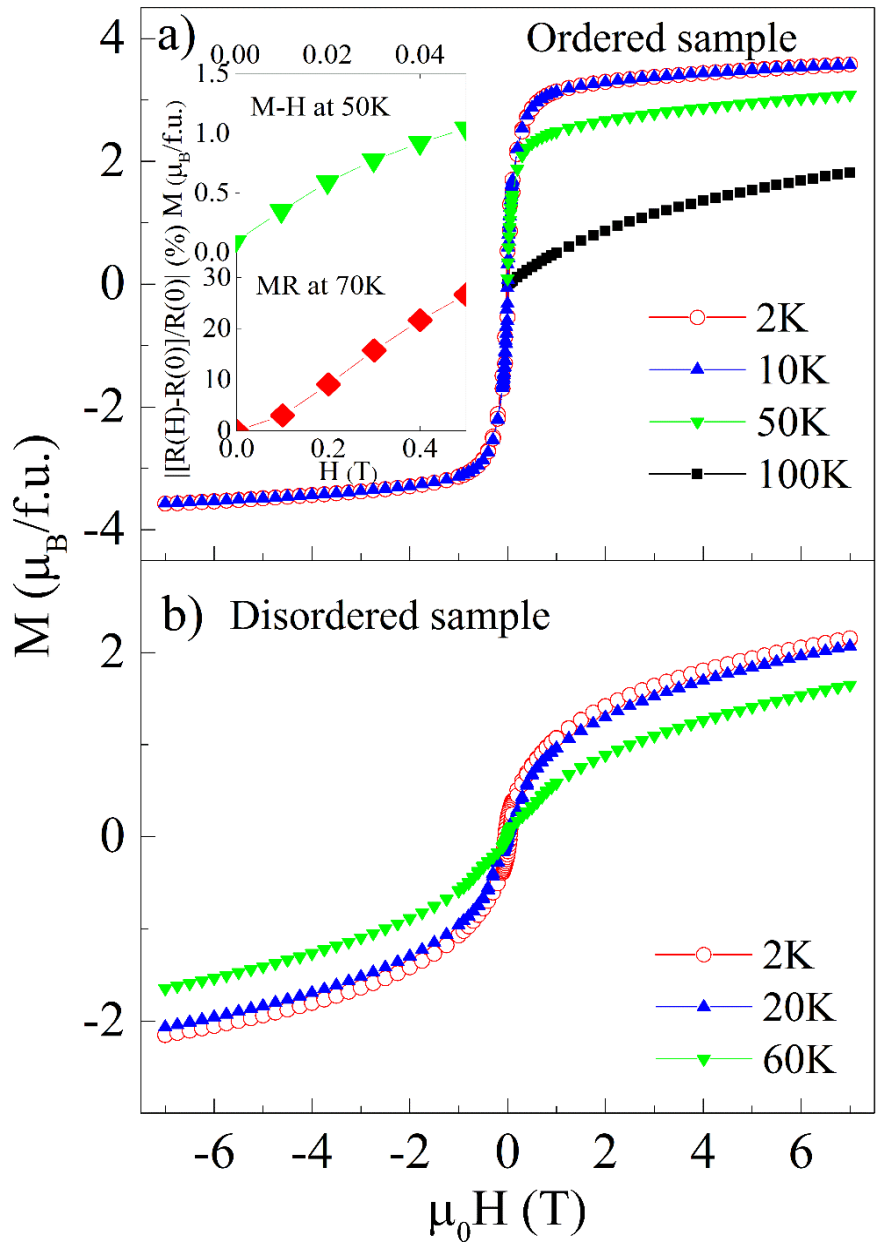

Figure 3. Magnetic hysteresis loops of ordered and disordered Tl$_2$NiMnO$_6$ at various temperatures. The inset in a) shows M-H curve at 50 K in the field range of 0-500 Oe and MR vs H at 70 K

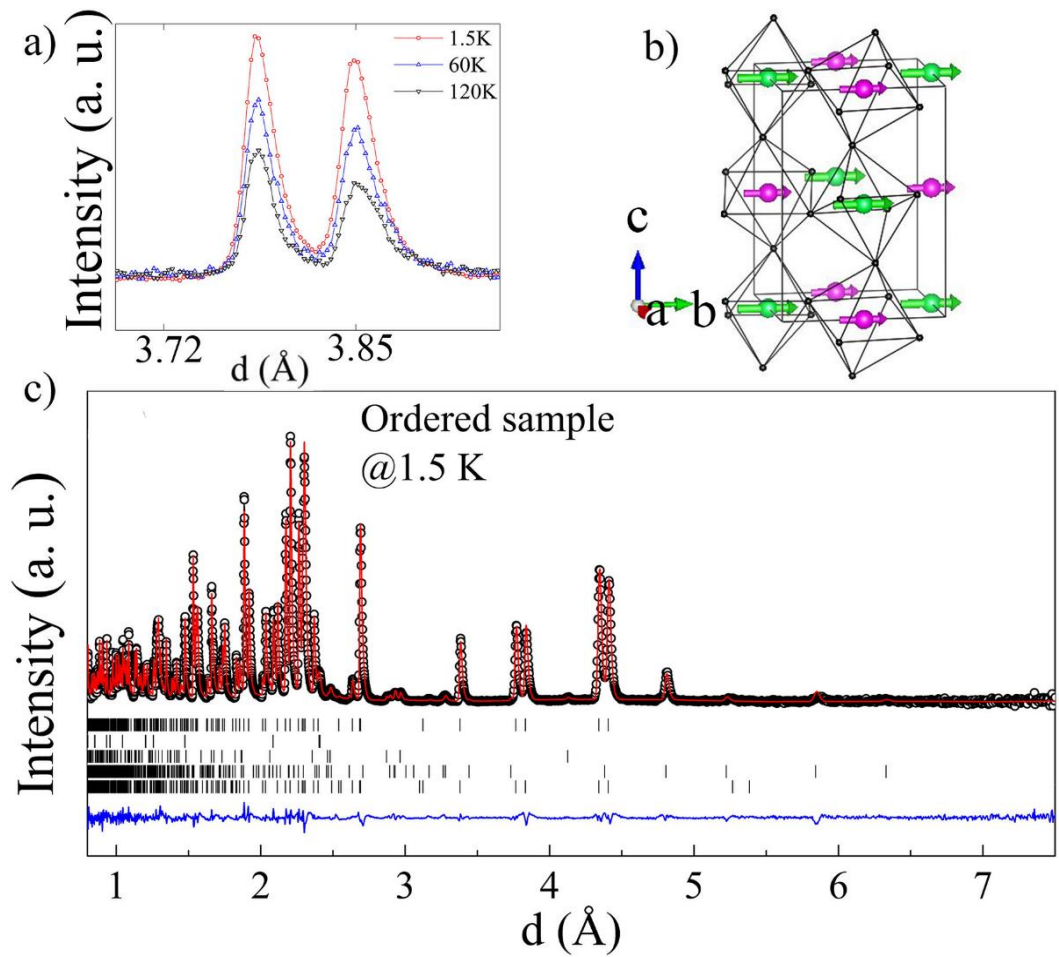

Figure 4. a) The 110 and 002 reflections, where the strongest magnetic contribution is observed, measured at different temperatures. b) Schematic representation of the magnetic structure for the ordered sample. c) Rietveld refinement of the neutron powder diffraction pattern of the ordered sample collected at 1.5 K. The nuclear reflections of $Tl_2NiMnO_6$ are denoted by upper tick marks. The lowest tick marks represent the magnetic reflections.

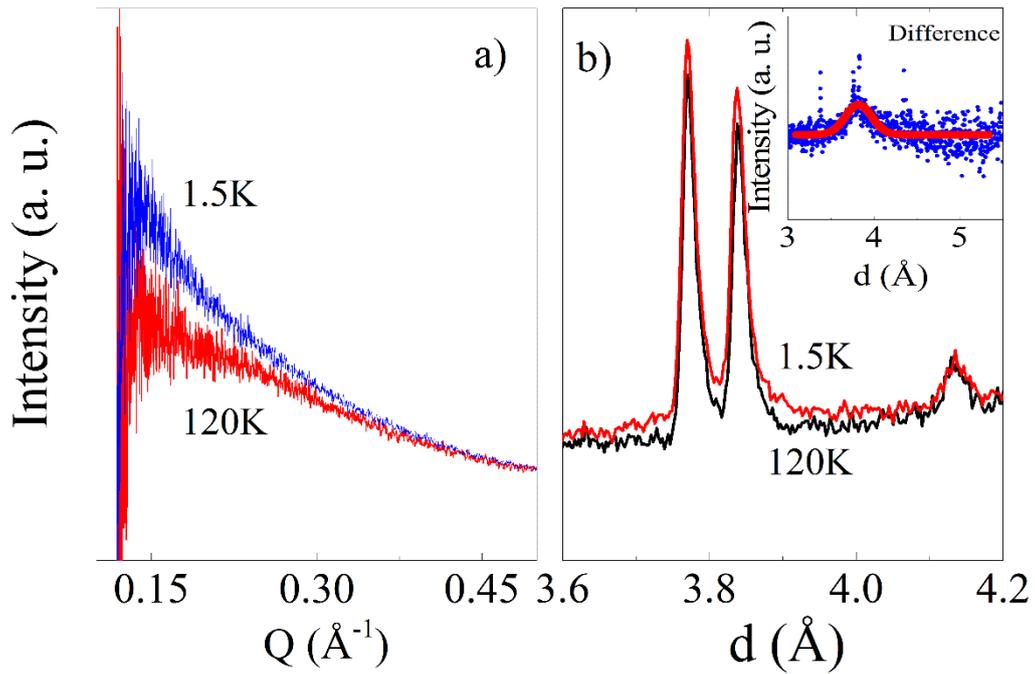

Figure 5. a) Neutron diffraction data for disordered $Tl_2NiMnO_6$ collected in forward scattering detector bank at 1.5 and 120 K. b) Reflections 110 and 002 at 1.5 and 120 K, highlighting the presence of weak diffuse scatting, centred at 3.82(1) Å from a profile fitting of the diffuse scattering data (inset).

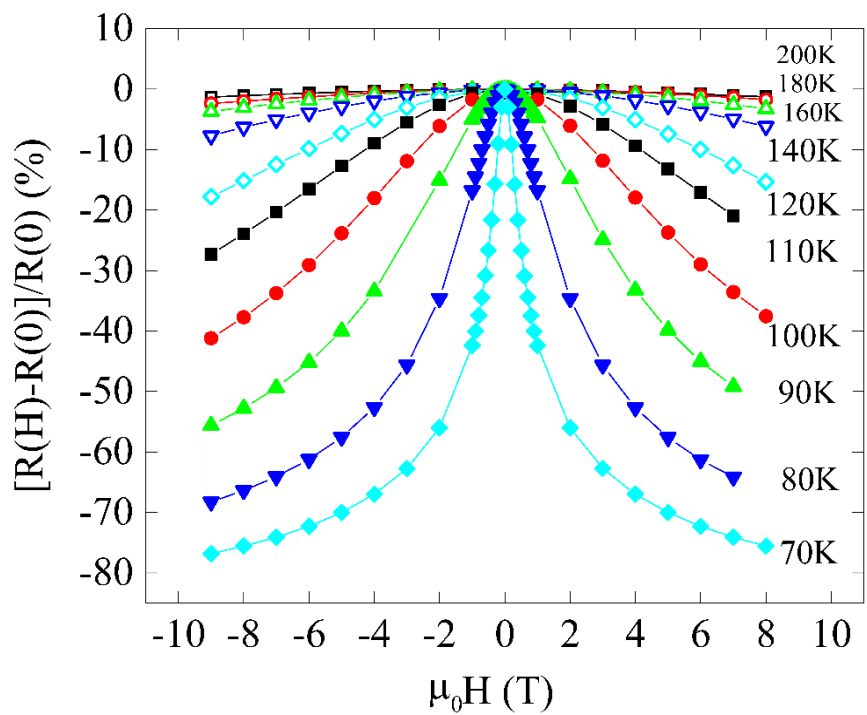

Figure 6. Isothermal magnetoresistance curves of ordered $Tl_2NiMnO_6$.

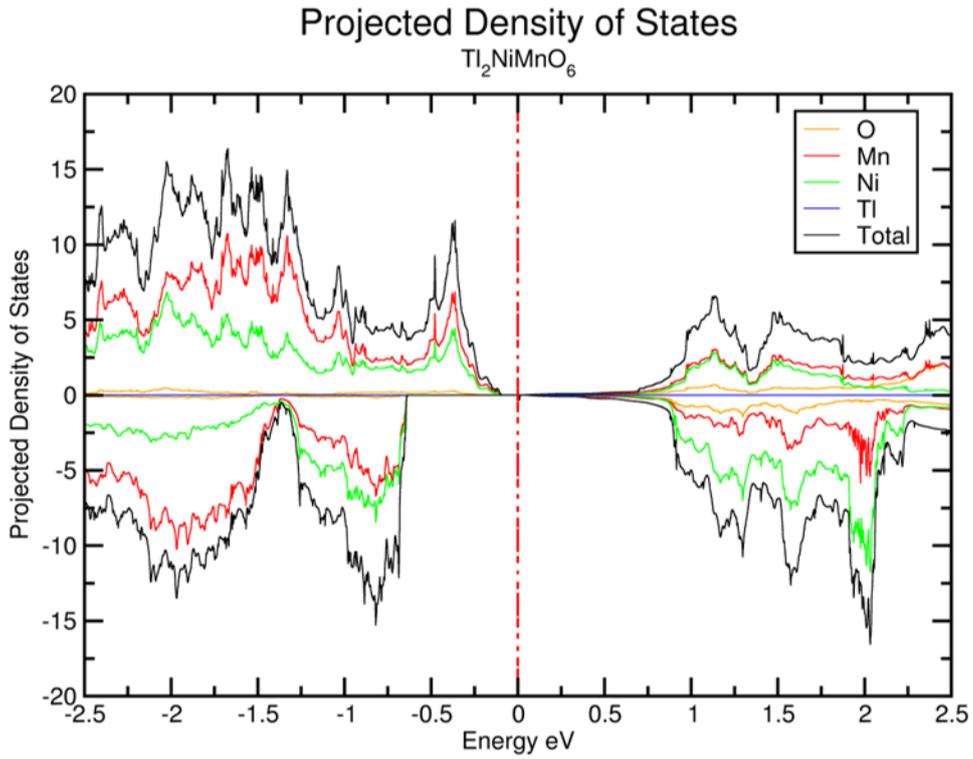

Figure 7. σ-DOS and PDOS of Tl$_2$NiMnO$_6$, where the Fermi Energy is the red dashed line and shifted to zero. Spin-polarisation is represented by α-electrons occupying the positive region of the PDOS-axis and β-electrons occupying the negative region.